\documentclass[12pt]{article}

\usepackage[english]{babel}

\usepackage[a4paper,top=3cm,bottom=3cm,left=3cm,right=3cm,marginparwidth=1.75cm]{geometry}
\usepackage[onehalfspacing]{setspace}
\usepackage{lineno} 

\usepackage{amsmath}
\usepackage{graphicx}
\usepackage[colorlinks=true, allcolors=blue]{hyperref}
\usepackage[square,numbers,sort&compress]{natbib}

\usepackage{authblk}
\title{Collective states of $\alpha$-sexithiophene chains inside boron nitride nanotubes}
\author[1,*]{Sabrina Juergensen}
\author[2]{Jean-Baptiste Marceau}
\author[1]{Chantal Mueller}
\author[3,4]{Eduardo B. Barros}
\author[1]{Patryk Kusch}
\author[1,5]{Antonio Setaro}
\author[2]{Etienne Gaufrès}
\author[1]{Stephanie Reich}

\affil[1]{Department of Physics, Freie Universität Berlin, Berlin, Germany. }
\affil[2]{Laboratoire Photonique Numérique et Nanosciences, Institut d’Optique, Université de Bordeaux, Bordeaux, France.}
\affil[3]{Department of Physics, Federal University of Ceará, Fortaleza, Ceará, Brazil.}
\affil[4]{Department of Physics, Technische Universität Berlin, Berlin, Germany.}
\affil[5]{Engineering Department, Pegaso University, Naples, Italy.}
\date{}                     
\begin{document}
\maketitle

\begin{abstract}
\noindent

Nanotubes align molecules into one dimensional chains creating collective  states through the coupling of the molecular transition dipole moments. These collective excitations have strong fluorescence, narrow bandwidth, and shifted emission/absorption energies. We study the optical properties of $\alpha$-sexithiophene chains in boron nitride nanotubes by combining fluorescence with far- and near-field absorption spectroscopy. The inner nanotube diameter determines the number of encapsulated molecular chains. A single  chain of $\alpha$-sexithiophene molecules has an optical absorption and emission spectrum that is red-shifted by almost 300\,meV compared to the monomer emission, which is much larger than expected from dipole-dipole coupling. The collective state splits into excitation and emission channels with a Stokes shift of 200\,meV for chains with two or more files. Our study emphasises the formation of a delocalized collective state through Coulomb coupling of the transition moments that shows a remarkable tuneability in transition energy. 

\end{abstract}

\newpage

\doublespacing

\section*{Introduction}
Collective states in one- and two-dimensional molecular aggregates are fascinating phenomena that indicate their presence by red-shifted, strong, and narrow light emission compared to the single molecule.~\cite{Bricks2017, Juergensen2023, Zhao2019} The collective state is created when the transition dipole moments of the single molecules couple to a macroscopic giant dipole that may exhibit superradiance, with potential applications in nanophotonics and optoelectronics. Collective states emerge when the molecules arrange spontaneously in a head-to-tail configuration (J-aggregate), but templating increases length and order  of a one-dimensional molecular aggregate. Perfect containers to align and string molecules into well-ordered one-dimensional chains and ensure their close packing are nanotubes with nanometer inner diameter.~\cite{Gaufres2013,Tunuguntla2016,Jordan2022,Hilder2009} In particular, boron nitride nanotubes (BNNTs) are ideal candidates to observe the strong fluorescence of  molecule chains. In contrast to the widely known carbon nanotubes (CNTs),~\cite{Wasserroth2019} BNNTs will keep the molecular emission intact due to the wide bandgap of boron nitride.~\cite{Allard2020} An added benefit of the nanotube environment is that it protects the fluorescent molecules against degradation and photobleaching extending the life span of the molecules.~\cite{Allard2020}

The inner tube diameter determines the number of molecular chains that may get encapsulated and aligned inside a nanotube. For small enough nanotube diameters, the alignment favors single molecular chains with a head-to-tail configuration. The exact diameter depends on the size of the nanotube, but typical diameters for single chains are $<1\,$nm.\cite{Badon2023} A one-dimensional chain will, \emph{e.g.}, align and add up the static dipole moments of molecules, which was shown to lead to macroscopic dipoles and a strongly non-linear optical response.~\cite{Cambre2015} It will also couple transient transition moments into collective molecular states.~\cite{Juergensen2023,Zhang2016,Deshmukh2019} The collective one-dimensional states manifest in a giant cross section in Raman scattering from molecules in carbon nanotubes (CNT) and an intense and strongly shifted fluorescence inside BNNTs.~\cite{Gaufres2013, Wasserroth2019, Allard2020} With increasing inner tube diameter, molecules form parallel lines or multi-file chains and, eventually, unordered molecular aggregates.~\cite{Gaufres2013,Allard2020,Gaufres2016,Badon2023} The number of parallel chains affect transition energies of the collective molecular states. For example, multi-file chains of $\alpha$-sexithiophene (6T) were observed to have distinct emission spectra from single-file chains.~\cite{Badon2023} Nevertheless, an understanding how the formation of J-aggregates and their side-by-side combination affects the collective molecular states in their absorption and emission behavior remains lacking at individual nanotube scale. 


Here, we show  how the arrangement of 6T molecules in BNNTs determines the properties of collective molecular states. We observed strongly red-shifted fluorescence and absorption peaks due to transitions in single- and multi-filed molecular chains. The overall red-shift of the optical transitions results from the J-type head-to-tail configuration inside the 6T chains. The $>10\%$ red shift of the energy of the collective state compared to the monomer cannot be explained by a point dipole description of molecular transition moments and even appears to exceed a reasonable description with extended transition dipoles. The side-by-side character in multi-file chains splits the absorption and emission lines as is typical in molecular H-aggregates. 
We complement far-field spectroscopy  with wavelength-tuneable scattering-type scanning near-field optical microscopy (s-SNOM) to analyse the 6T@BNNT complex with a spatial resolution of 20\,nm and resolve absorption along different parts of the BNNT to optically separate the molecular lattices in encapsulation.

\section*{Sample Characterization}

The BNNTs with encapsulated dye (6T@BNNT) were deposited onto a Si/SiO$_2$ wafer by spin coating. Encapsulation and deposition are described in the Methods. The obtained sample was characterised by combined AFM and spatial fluorescence spectroscopy as well as polarization-dependent spatial modulation spectroscopy.

\begin{figure}[h]
\centering
\includegraphics[width=0.95\textwidth]{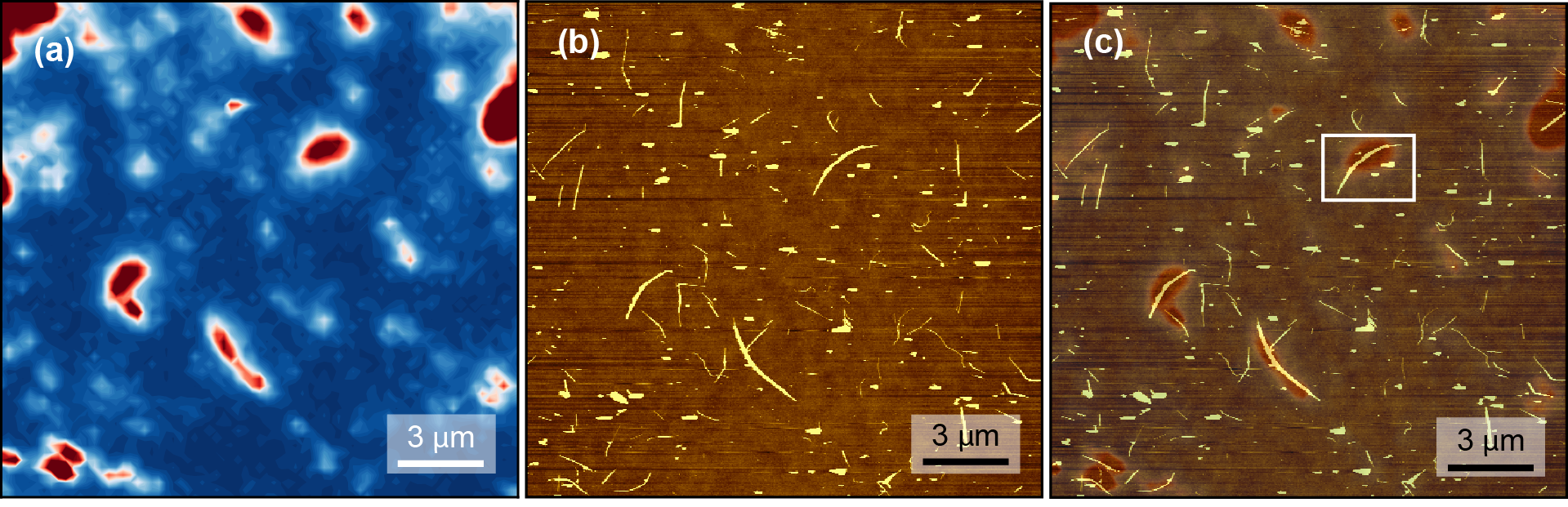}
\caption{\label{fig:AFM_PL} \textbf{Fluorescence spectroscopy and atomic force microscopy of 6T@BNNT.} \textbf{(a)} Spatial fluorescence map of 6T@BNNT on Si/SiO$_2$. Showing the fluorescence of the main transition (2.10-2.14\,eV) of the 6T molecules. Red areas show the highest intensity. \textbf{(b)} AFM topography image of the sample area in panel a. \textbf{(c)} Fluorescence map on top of the AFM image to identify tubes are filled with 6T. The white rectangle marks the tube that was studied in the following.}
\end{figure}

Fluorescence microscopy was combined with AFM to distinguish between filled and empty BNNTs as well as between filled tubes and molecular agglomerates.
A spatial fluorescence map of a 16~$\times$~16\,$\mu m$ sample area was recorded, Fig.~\ref{fig:AFM_PL}a, using 2.33\,eV (532\,nm) laser excitation, which is close to the molecular resonance.~\cite{Wasserroth2019} It maps the intensity of the strongest fluorescence peak of 6T over the coordinates of the scanned sample area showing areas of strong fluorescence (red areas) that may originate from filled tubes or free molecular agglomerates on the substrate. To distinguish between agglomerates and filled BNNTs we combine the fluorescence map with an AFM image of the same sample area, see Fig.~\ref{fig:AFM_PL}b. Laying the fluorescence map on top of the AFM image, Fig.~\ref{fig:AFM_PL}c, shows that the majority of the fluorescent spots indeed come from filled BNNTs. We identified a strongly emitting 6T@BNNT - likely a bundle of filled tubes - which is marked by a rectangle in Fig.~\ref{fig:AFM_PL}c. The following measurements were performed on this 6T@BNNT unless noted otherwise. There are also empty tubes without molecules that do not show up in the fluorescence map, but are present in the AFM image.

\begin{figure}[h]
\centering
\includegraphics[width=0.95\textwidth]{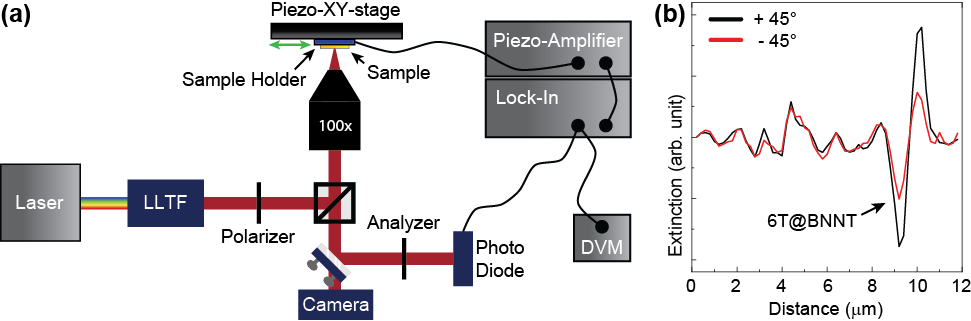}
\caption{\label{fig:SMS_Pol} \textbf{Spatial modulation spectroscopy.} \textbf{(a)} Sketch of the spatial modulation setup. \textbf{(b)} Polarisation-dependent extinction line profile of the in SI Fig.~\ref{fig:SI_SMS_Map} marked tube taken at 530\,nm excitation wavelength. Black line +45° and red line -45° polarized light.}
\end{figure}

To analyse molecular alignment in the filled BNNTs we measured the extinction of 6T@BNNT, \emph{i.e.}, the sum of absorption and scattering, using polarized SMS. A sketch of the SMS setup is depicted in Fig.~\ref{fig:SMS_Pol}a, see Methods for details. The sample holder contains a piezo element that is mounted to modulate the sample position with frequency~$f$ while the sample is scanned through the focus of a laser beam. The demodulation of the scattered signal with a lock-in amplifier at frequency~$f$ leads to a contour plot with the shape of the first derivative of a Gaussian profile, from which we determine the extinction cross section of the tube bundle.~\cite{Arbouet2004,Devadas2014,Devadas2015} 

We first recorded polarisation-dependent SMS using 2.34\,eV (530\,nm) laser excitation, Fig.~\ref{fig:SMS_Pol}b and Supplementary Fig.~\ref{fig:SI_SMS_Map}a. It highlights the polarisation dependence of the tube bundle. When the light is polarized along the BNNT axis the extinction is maximised (black line in Fig.~\ref{fig:SMS_Pol}b) and when the polarization is (almost) orthogonal to the tube it is minimal (red line). Other objects like dirt and non ordered molecular agglomerates do not show any polarisation dependence, see Supplementary Fig.~\ref{fig:SI_SMS_Map}a. This agrees with the prior observations of the polarisation dependence of 6T molecules inside BNNTs~\cite{Badon2023} and shows that the 6T molecules are aligned along the tube axis. The polarization-dependent SMS measurements also allowed us to identify molecular agglomerates outside BNNTS, see Supplementary Fig.~\ref{fig:SI_SMS_Map}a. From the 6T@BNNT characterization we thus identified 6T@BNNTs, unfilled BNNTs, and 6T agglomerates and dirt outside of BNNTs. 

\section*{Results}
\begin{figure}[h]
\centering
\includegraphics[width=0.98\textwidth]{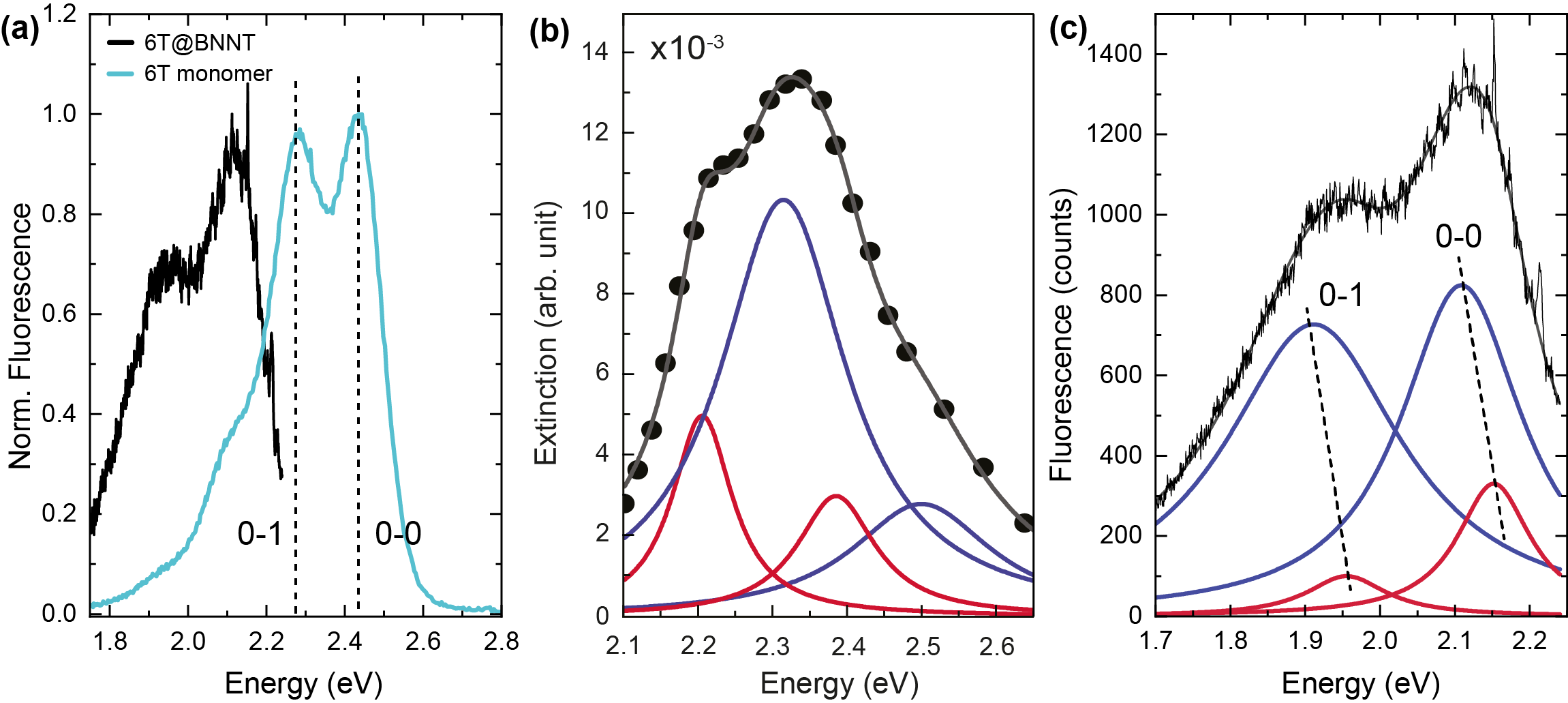}
\caption{\label{fig:PL} \textbf{Fluorescence Spectra and Wavelength-Dependent Spatial Modulation Spectroscopy.} \textbf{(a)} The fluorescence spectrum of 6T@BNNT (black) and 6T monomer (light blue) show a frequency shift indicating the formation of ordered molecular aggregates inside the tube. \textbf{(b)} Wavelength-dependent extinction profile of the 6T@BNNT bundle. \textbf{(c)} Fluorescence spectrum of 6T@BNNT (black) fitted by four Lorentzian fits (red and blue solid lines). Dots are measurements and solid lines are Lorentzian fits.}
\end{figure}

A 6T@BNNT fluorescence spectrum is plotted in Fig.~\ref{fig:PL}a together with the monomer spectrum of 6T measured in solution. In the monomer, the two main peaks were assigned to the 0-0 (2.44\,eV) zero-phonon line and its 0-1 (2.29\,eV) phonon replica with an energy separation of 0.15\,eV~\cite{Zhao2021,Bhagat2021,Milder2009}. The emission of the 6T@BNNT  is strongly shifted to smaller energies as  is typical for the formation of molecular aggregates.~\cite{Sun2020,Deshmukh2022} In addition, the energetic separation of the two main peaks increased for the encapsulated molecules compared to the monomer, Fig.~\ref{fig:PL}a and Table~\ref{tab:energie}, and the intensity ratio changed.



To study the absorption of the 6T@BNNTs we performed wavelength-dependent SMS on the same sample area, giving us the extinction profile of the 6T@BNNT bundle in Fig.~\ref{fig:PL}b. It is noticeable that the extinction is not the mirror image of the emission spectra in Fig.~\ref{fig:PL}a, as is typical for molecules due to the similar emission/absorption probabilities in the ground (S$_0$) and first excited electronic state (S$_1$).~\cite{Lakowicz,Valeur2012} The energetic separation of the two main peaks in the extinction spectrum ($\Delta\omega$\,=\,0.13\,eV) differs from the 6T@BNNT and monomer fluorescence spectra in Fig.~\ref{fig:PL}a, see also Table~\ref{tab:energie}. 

The apparent discrepancies in emission and extinction spectrum of the 6T@BNNT arise from tubes filled with different numbers of parallel  molecular chains resulting in a superposition of their optical responses.~\cite{Badon2023} The fluorescence spectrum in Fig.~\ref{fig:PL}c is fitted well by four Lorentzians that we interpret as two pairs of peaks (red and blue lines). We assign them to the zero-phonon line and the phonon replica of the collective molecular states in single- and multi-file chains. Up to three chains may lie next to each other in ordered aggregates in a tube leading to a head-to-tail configuration of the transition dipoles along the tube axis and a side-by-side configuration perpendicular to the axis.~\cite{Badon2023} For tubes with an inner diameter of more than 2.5\,nm the molecules start to form undefined aggregates in the tube leading to an optical response that was not observed in the spectra presented here.~\cite{Song2023} 
The fitting of the extinction curve with four Lorentzians, see Fig.~\ref{fig:PL}b, results in two extinction spectra (red and blue) that are the expected mirror images of their respective 6T@BNNT emission, Fig.~\ref{fig:PL}c. The energy difference of the two transitions fits to the C=C stretching vibration ($\sim$\,0.18\,eV), see Table~\ref{tab:energie}.~\cite{Blinov1995,Kouki2000,Zhao2021} Despite the similar appearance, the double peak structure of the emission and extinction spectra have different origin: The two peaks in the extinction spectrum originate from the 0-0 transitions of the single and multi-file chains, while the fluorescence is dominated by the 0-0 and 0-1 multi-file emission, Fig.~\ref{fig:PL}c.

\begin{table}[h]
   \centering
   \caption{Transition energies and linewidth of the individual and collective 6T excitons.}
   \vspace{0.4cm}
   \begin{tabular}{lcccc}
   \hline
& \multicolumn{2}{c}{$S_{0-0}$} & \multicolumn{2}{c}{$S_{0-1}$}  \\
\hline
& $\omega$ (eV) & FWHM (eV)&  $\omega$ (eV) & FWHM (eV) \\
\hline
\multicolumn{5}{c}{\textbf{monomer}}\\
\hline
emission  &   2.44 &  0.14 & 2.29& 0.15\\
\hline
\multicolumn{5}{c}{\textbf{single file}}\\
\hline
emission &  2.15  & 0.12 & 1.96& 0.13\\
extinction &  2.21 &  0.10 & 2.39 & 0.13\\
absorption & 2.16  & 0.14 & 2.25 - 2.40 & - \\
\hline
\multicolumn{5}{c}{\textbf{multi-file}}\\
\hline
emission &  2.11 & 0.20  & 1.91& 0.29\\
extinction&  2.31 & 0.21 & 2.50 & 0.23\\
\hline
    \end{tabular}
    \label{tab:energie}
\end{table}

\begin{figure}[h]
\centering
\includegraphics[width=1\textwidth]{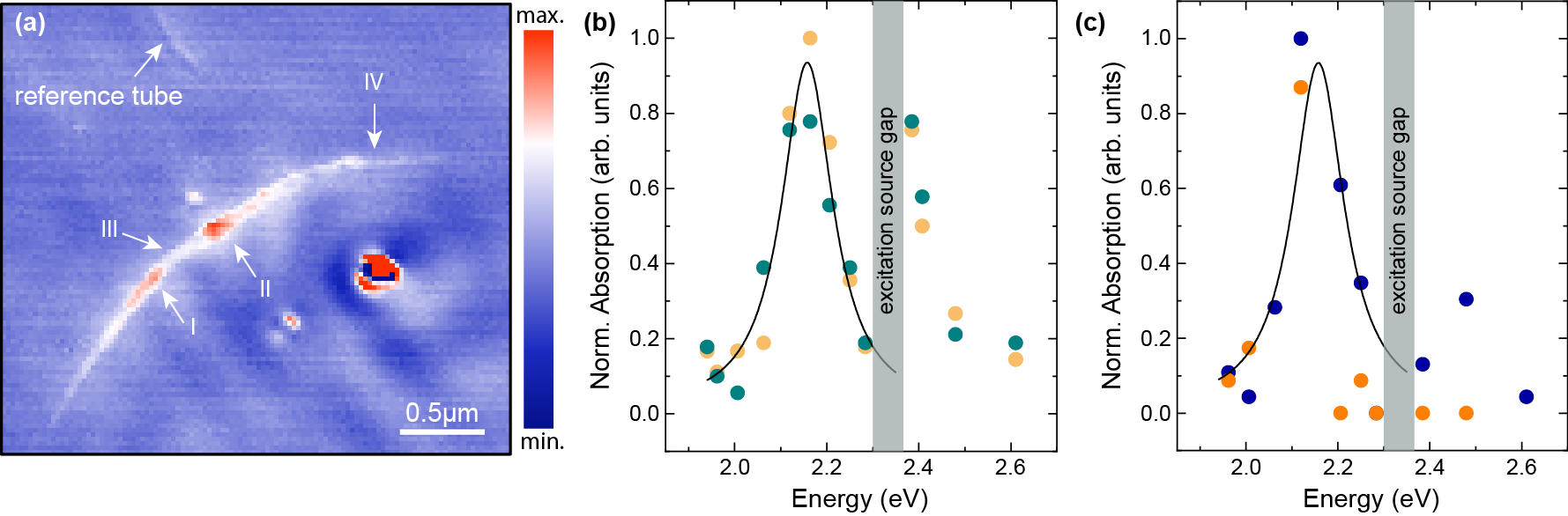}
\caption{\label{fig:SNOM} \textbf{s-SNOM measurements of the 6T@BNNT bundle.} \textbf{(a)} s-SNOM phase image of a 6T@BNNT hybrid. The red areas indicate a higher absorption than the white and blue areas. \textbf{(b)} s-SNOM measurements that belong to two parts of the BNNT bundle in (a) with higher absorption. The yellow (I) and petrol blue (II) curve. The solid line is the Lorentzian fit of the absorbance. \textbf{(c)} The dark blue (III) and orange (IV) curve belong to two parts of the BNNT bundle in (a) that absorb less. For better comparison the absorption spectra are normalized to unity in panels b and c.}
\end{figure}

We used scattering scanning near-field optical microscopy (s-SNOM) to study the 6T@BNNT absorbance and its distribution along the nanotube axis with a resolution of $\sim$\,20\,nm. s-SNOM determines the local absorbance of nanomaterials from its phase images, see Methods,~\cite{Taubner2004,Mastel2015,Niehues2023} and allows to study various parts and hotspots along the 6T@BNNT bundle. Using a wavelength tuneable laser we measure phase images of the same 6T@BNNT bundle as a function of energy. An exemplary phase image recorded at 2.25\,eV (550\,nm) excitation is shown in Fig.~\ref{fig:SNOM}a. The red areas indicate higher absorption than the white and blue areas. Figure~\ref{fig:SNOM}b,c show the calibrated phase signal (absorbance) over 1.9\,-\,2.7\,eV excitation energy at four different positions along the tube, arrows in Fig.~\ref{fig:SNOM}a.

All four positions (I-IV) have an absorption peak at 2.16\,eV. Positions I and II may show a second absorption maximum between 2.25\,-\,2.40\,eV, but an energy gap in the tunable laser prevents measurements in this range. We identify the absorption peak at 2.16\,eV in Fig.~\ref{fig:SNOM}b,c with the single file transition at 2.21\,eV in the extinction spectrum. The lowest transition of the extinction profile is thus by 50\,meV higher than the peak in the absorption spectrum measured with the s-SNOM, see Table~\ref{tab:energie}. This shift may be due to a strong scattering contribution to the extinction data and/or the characteristic shift between optical near- and far-field measurements.~\cite{Zuloaga2011,Wasserroth2018,EstradaReal2023} We also note that the near-field absorption energy is identical to the far-field emission, which would point towards a vanishing Stokes shift for the single file. The variation of the absorbance along the tube bundle indicated the tubes may be only partially filled or that the tube bundle gets thinner towards its end, since the s-SNOM signal is proportional the the amount of material. This also agrees with the observation that spots III and IV appear to have no absorption associated with the multi-file chains.

\section*{Discussion}

\begin{figure}[h]
\centering
\includegraphics[width=0.98\textwidth]{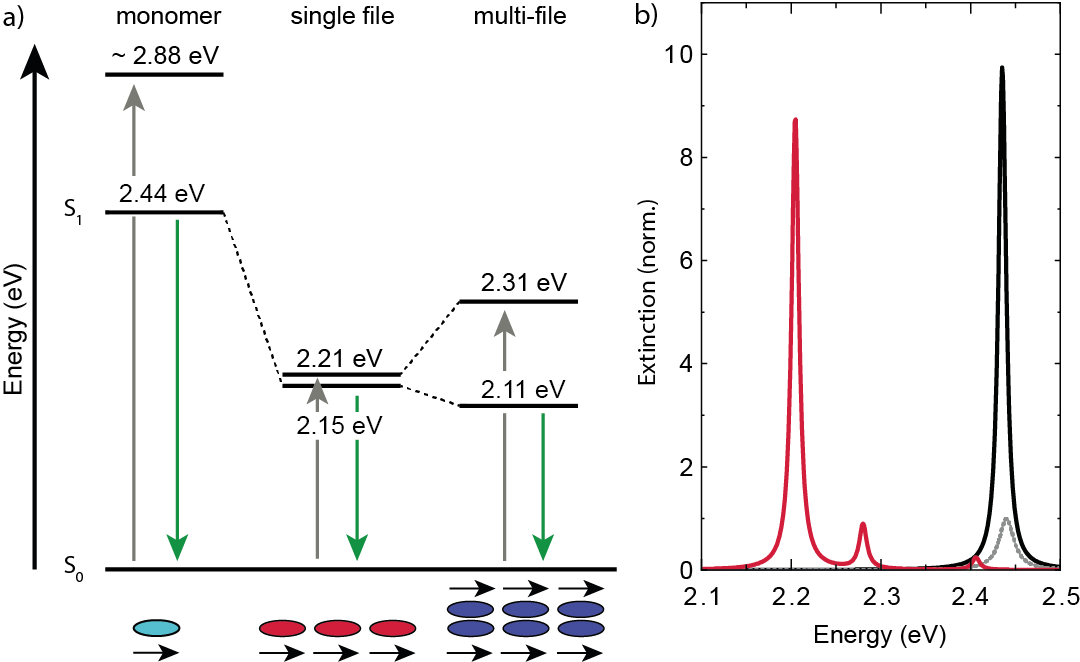}
\caption{\label{fig:Jablonski} \textbf{a) Jablonski Diagram of the 6T@BNNT Bundle and Monomer.} Illustration of the absorption (grey arrows) and fluorescence (green arrows) measurements. Showing that the encapsulated molecules show a red shift of the absorption and emission. Indicating the formation of agglomerates with strong J-character. \textbf{b) Calculations of the Extinction of a 6T Single File Chain.} Calculation without (black) and with (red) correction factor $g = 2.422$\,nm. Both spectra are normalized to the extinction of a single dipole (gray dashed) with $\omega_0 = 2.44\,$eV, $d_0=8.45$\,D, and $\gamma_0=5\,$meV. Further parameters: lattice constant $a= 2.51\,$nm, $\epsilon_m=2.55$, and number of molecules $N = 10$. The dipole polarization was chosen along the $x$ direction.}
\end{figure}

The fluorescence and absorption spectrum of the 6T@BNNTs reveal a rich combination of excited collective states as pictured in the Jablonski diagram in Fig.~\ref{fig:Jablonski}, see also Table~\ref{tab:energie}. The diagram compares the energy of the first excited state of the 6T monomer, see Supplementary Fig.~\ref{fig:Comparison}, with the collective states in 6T@BNNT single- and multi-file chains. The red shift of the collective state in the single 6T chain comes from the pure head-to-tail character along the chain, leading to a collective lowest excited states with an allowed dipole transition. The fluorescence of the single chain is also red shifted as is typical for J-aggregates.~\cite{Bricks2017, Spano2009} The remaining Stokes shift of 60\,meV is vanishingly small compared to the $\sim$\,440\,meV shift of the monomer, which also indicates a pure head-to-tail configuration.

The collective energies of the 6T in BNNT agrees excellently with previous measurements of 6T transitions in carbon nanotubes performed by resonant Raman scattering.~\cite{Wasserroth2019} The close similarity of the carbon and boron nitride encapsulated implies that  the packing of the 6T is similar in the two types of nanotubes and governed by the geometry of the available space. The difference in dielectric environment of carbon versus boron nitrite nanotube plays a negligible role for the energetic position of the molecular aggregates, which is rather surprising but agrees well with the identical transition energies of ordered dye monolayers on two-dimensional hBN and graphene.~\cite{Juergensen2023} It is also confirmed by the weak dielectric screening by semiconducting nanotubes measured in double-walled nanotubes.~\cite{Gordeev2024}

The collective shift of the 6T chains (290\,meV) is very large compared to the observed red shifts in 2D molecular lattices (60\,meV).~\cite{Zhao2019, Juergensen2023} Indeed, simulations of the 6T chains as a string of point dipoles separated by the molecular distance (2.51\,nm) predict only a shift of 5\,meV, black and dashed gray spectra in Fig.~\ref{fig:Jablonski}b, although this same model worked excellently for 2D lattices.~\cite{Juergensen2023} The applied model can be extended to take into consideration the finite size of the molecules,~\cite{Romaner2008} yielding a much better fit to the experimental data, red line in Fig.~\ref{fig:Jablonski}b. However, the necessary correction amounts to an effective dipole displacement on the same order of magnitude (96.5\,\%) as the physical length of 6T, which appears illusive. The physical origin of the extremely strong molecule-molecule interaction inside the nanotube remains unreasonable, and should be the subject of further inquiry. One possibility is a waveguiding effect of the encapsulating tube that enhances the Coulomb interaction between the molecular transition dipoles.


The multi-file chain shows a smaller red shift of the absorption and a slightly larger red shift of the fluorescence, \emph{i.e.}, effectively a splitting of the collective single-file state, Fig.~\ref{fig:Jablonski}a. The side-by-side configuration of the chains adds an anti-bonding configuration to the coupling of transition moments resulting in the splitting. The closer packing of the molecular chains compared to the in-chain separation leads to a stronger splitting of the excited state and thus an emission that is shifted further to the red compared to the single chain and in a larger Stokes shift (200\,meV).~\cite{Spano2009,Gierschner2013, Eder2022} The  shift of the absorption peak to smaller energies compared to the 6T monomer verifies that the J-character dominates the optical properties as expected for a  multi-file molecular chain. Chains with a side-by-side configuration, \emph{i.e.}, the molecules are oriented perpendicular to the nanotube axis, would result in a collective absorption at higher energies than the monomer and a further increase in Stokes shift,~\cite{Bricks2017, Spano2009} in contrast to the dual red shift and smaller Stokes shift observed experimentally.

The full width at half maximum (FWHM) of the 6T@BNNT spectrum further confirms our assignment of single- and multi-file chains. The single-chain emission is narrower than the monomer and the multi-file emission, see Fig.~\ref{fig:PL}a,c and Table~\ref{tab:energie}. The collective state of J-aggregates has a reduced inhomogeneous broadening explaining the narrowing compared to the monomer.~\cite{Knapp1984,Juergensen2023} In addition, the absorption of the single chain is narrower than for several 6T files, because J-aggregates absorb predominantly into the lowest vibronic state whereas H-aggregates absorb in different vibronic levels of the excited state leading to a broadening of the absorption features.~\cite{Sauer2011,Juergensen2023}

Finally, we comment on the relative intensities of the single- and multi-file chains in fluorescence. J-aggregates are expected to show brighter emission from their collective state than H aggregates,~\cite{Juergensen2023,Sun2020,Eisfeld2017,Marceau2024} but we observed stronger from the multi-file chains. The observed intensities are most likely dominated by the smaller amount of tubes with single than with multi-file chains.~\cite{Badon2023,Marceau2024} This is consistent with our observations of the extinction. The 0-0 emission is, in fact, dipole-forbidden for H-aggregates. Therefore, the peak intensity of the 0-0 emission decreases with increasing number of 6T chains inside the tube and vanishes for pure H-aggregates.~\cite{Loi2004, Yamagata2012, Bhagat2021, Eder2022}

\section*{Conclusion}
In conclusion we studied absorption and emission from collective optical states of encapsulated 6T chains in BNNTs. By comparing the energetic position of the fluorescence and absorption the Stokes shift allowed us to identify the formation of single- and multi-file chains inside the nanotubes. The dominant configuration for the energy of the collective state is the J-type head-to-tail arrangement, we did not observe chains where the molecules are aligned in pure H-configuration or randomly orientated inside the tubes. The use of s-SNOM with its high spatial resolution made it possible to resolve which part of the tube bundle has a stronger absorption than other parts which is not possible with standard microscopic methods due to the large laser spot size covering the whole nanostructure. The very large shift of the single-file chain could not be modeled by the microscopic dipole model without the correction of the finite molecule size, requiring a further theoretical investigation.

\section*{Methods}

\subsection*{Sample Preparation}
\textbf{6T@BNNT:} Boron nitride nanotubes (BNNTs) were provided by BNNT Materials and the Canadian National Research Center. $\alpha$-sexithiophene (6T), toluene, and N,N-dimethylformamide (DMF) were purchased from Sigma-Aldrich and used as received. Thereby, only reagent grade solvents were used.

For cleaning and opening the BNNT powder was first annealed for 2 hours at 800$^{\circ}$C under ambient conditions and afterwards sonicated in DMF using a cup sonicator until complete dispersion was observed. In the next step, the BNNT solution was centrifuged at 12000g and the top half of the centrifuge tube was collected and filtered by a polytetrafluoroethylene (PTFE) membrane with a pore size of 0.22\,$\mu$m. The encapsulation was done under reflux for 48h at 85$^{\circ}$C by mixing 6T molecules and the cleaned BNNTs with toluene in a round-bottom flask equipped with a condenser. Thereby, the concentration of the 6T was fixed at $5 \times 10^{-6}\,$M. Intense DMF rinsing of the solution on a PTFE membrane (pore size 0.22\,$\mu$m) was applied to remove non-encapsulated 6T molecules. Afterwards soft piranha treatment was done to remove the last free 6Ts and the 6Ts that were adsorbed on the outer wall of the BNNTs.

To dispose the 6T@BNNTs \textit{via} spin coating onto a Si/SiO$_2$ substrate, the 6T@BNNTs were dispersed in DMF. Before disposition the Si/SiO$_2$ substrate was sonicated 10\,min in acetone and 10\,min in isopropanol for cleaning. \\

\textbf{6T solution:} The 12\,$\mu$M solution for the emission and absorption measurements was prepared by dissolving $\alpha$-sexithiophene purchased by Sigma Aldrich in toluene ($\geq$ 99.7\%) form Honeywell.

\subsection*{Scattering Nearfield Optical Microscope (SNOM)}
The s-SNOM measurements were done with a commercial NeaSNOM from neaspec/Attocube in pseudoheterodyne configuration for background subtraction. As a probe we used platinum coated AFM tips from Nano World (Arrow-NCPt). As an excitation light source a wavelength-tuneable laser from Hübner (C-Wave) was used, covering the visible spectral range from 450 to 650\,nm.

The laser beam was guided through a beam expander onto a parabolic mirror (NA\,0.4) that focuses the beam onto the AFM tip which acts as a near-field probe in the visible spectral range. The incoming laser light creates a strong localized electromagnetic near field at the tip apex. When the tip is in the proximity of the sample the near field interacts with the sample changing the tip-scattered field. The detected light is composed of near-field signal from sample and background signal.

To extract the near-field signal from the background signal the tip is oscillated at a certain characteristic frequency $\Omega$ (300\,kHz). For further noise reduction a reference beam is send to a modified Michelson interferometer with oscillating mirror at frequency $M \ll \Omega$. The detected interfering beam is modulated at frequencies $f=n\Omega \pm mM$ leading to side bands next to the fundamental harmonic (n$\Omega$).~\cite{Ocelic2006} Here $n$ and $m$ are integers expressing the order of the modulation. Demodulation suppresses the background signal and extracts the phase and the amplitude of the elastically scattered light. Thereby, the phase is proportional to the absorption of the investigated material and the amplitude to the reflectivity. Calibration of the 6T@BNNT phase signal was done by an empty BNNT to account for wavelength-dependent changes in the sensitivity of the setup. 

\subsection*{Spatial Modulation Spectroscopy}
The extinction was measured with a home build spatial modulation setup. As excitation source an supercontinuum laser (NKT - FIU-15) was used. The laser was coupled to an laser line tunable filter (Photon etc. - LLTF) to filter out single wavelengths. The laser beam was guided to an inverse microscope that was equipped with a $xy$-piezostage form PI and focused with an 100x objective (NA 0.9) to the sample. The sample was mounted in a special sample holder that was equipped with a piezo-electric element to modulate the sample position at frequency $f$ in the focal spot of the objective. The reflected signal was again collected by the 100x objective and detected with a switchable gain detector (Thorlabs - PDA36A2) that was connected to a lock-in amplifier (AMETEK - Model 7270) to demodulate the signal and further to a digital voltmeter (NI - myDAQ) to measure the total reflection. Yielding to a relative change in the reflectance $\Delta R / R$.

For the polarisation dependent measurements a polariser was insert into the beam path before the sample and an analyser was added in front of the detector. The analyser was kept constant to avoid changes in the signal intensity due to the sensitivity of the detector while the polariser was rotated.


\subsection*{Atomic Force Microscpy (AFM)}
The AFM images were taken with an XE-150 AFM from ParkSystems in non-contact mode. As a probe a silicon SSS-NCHR AFM tip from Nanosensor was used. 

\subsection*{Fluorescence \& Photoluminescence Excitation Spectroscopy }
The fluorescence measurements of the 6T@BNNT on silicon were performed with a XploRA (Horiba) Raman spectrometer which is equipped with an $xy$-piezostage from PI. As excitation wavelength a 532\,nm green laser diode was used. The spatial fluorescence maps were taken with an laser power of 0.42\,mW and the integration time was 0.5\,s. The step size of the piezostage was 250\,nm in $x$ and $y$ direction.\\

The fluorescence of the 6T solution was measured with HORIBA Jobin Yvon Fluorolog-3 spectrofluorometer using a Xe-lamp for illumination. The spectrofluorometer contains a double grating monochromator to select an excitation wavelength and an spectrometer to which an CCD is coupled which can detect wavelengths between 200\,nm and 850\,nm.

\subsection*{Microscopic Model of Collective Dipoles.}
To describe the molecule chains theoretically we use the dipole model of Ref.~\cite{Juergensen2023}. The eletric field influenced by the surrounding dipoles is calculated by the Greens function
\begin{equation}
	\begin{split}
		\mathrm{G}(\mathbf{r}_{ij})\mathbf{d}_j =& \frac{k^3}{4\pi\epsilon_0\epsilon_m}\mathrm{e}^{\mathrm{i}k r_{ij}} \Bigg[\left( \frac{1}{k r_{ij}} + \frac{\mathrm{i}}{(k r_{ij})^2} - \frac{1}{(k r_{ij})^3} \right)\mathbf{d}_j  
		\\& - \left( \frac{1}{k r_{ij}} + \frac{3\mathrm{i}}{(k r_{ij})^2} - \frac{3}{(k r_{ij})^3} \right) (\hat{\mathbf{r}}_{ij}\cdot \mathbf{d}_j)\hat{\mathbf{r}}_{ij} \Bigg],
	\end{split}
 \label{eq:greensfunction}
\end{equation}
where $\mathbf{r}_{ij} = \mathbf{r}_i - \mathbf{r}_j$, $r_{ij}=\vert \mathbf{r}_{ij} \vert$ represents the distance between the interacting molecules, $\mathbf{d}_j$ the individual dipole moments, and $\epsilon_m$ is the dielectric screening by the surroundings. To investigate the influence of the breakdown of the point dipole approximation, we modified the equation by adding a correction that accounts for the finite size of the molecule. In principle, the finite dipole size should affect each of the terms in Eq~\ref{eq:greensfunction} differently. However, for simplicity, we chose to simply substitute $kr_{ij}$ in all terms by $k\sqrt{r_{ij}^2-g^2}$ where $g$ is the same constant for all terms.\cite{Romaner2008} The dipole-dipole interaction in this system is dominated by the terms proportional to $1/(kr_{ij})^3$, which will be similarly affected by this correction, justifying the use of a single correction parameter.


\section*{Acknowledgements}
This work was supported by the European Research Council (ERC) under grant DarkSERS-772\,108, the German Science Foundation (DFG) under grant Re2654/13 and Re2644/10, and the SupraFAB Research Center at Freie Universit\"at Berlin. The authors thank BNNT LLC and C. Kingston from NRC Canada for providing raw BNNTs. EG acknowledges funding from CNRS starting package and the CNRS Tremplin program. PK acknowledges the DFG for funding (KU4034 2-1). AS gratefully acknowledges the financial support of the Italian Ministry of University and Research (MUR), Research Grant PRIN PNR 2022 No. DRASTIC, and Pegaso University within the PRA ed FRC funding scheme, grants BIOMAPS and ORIONE.

\bibliographystyle{naturemag}
\bibliography{6T-BNNT}

\newpage

\section*{Supplementary}

\renewcommand{\figurename}{Supplementary Fig.}
\setcounter{figure}{0}

\section{Polarisation dependent spatial modulation maps}
\begin{figure}[h]
\centering
\includegraphics[width=0.9\textwidth]{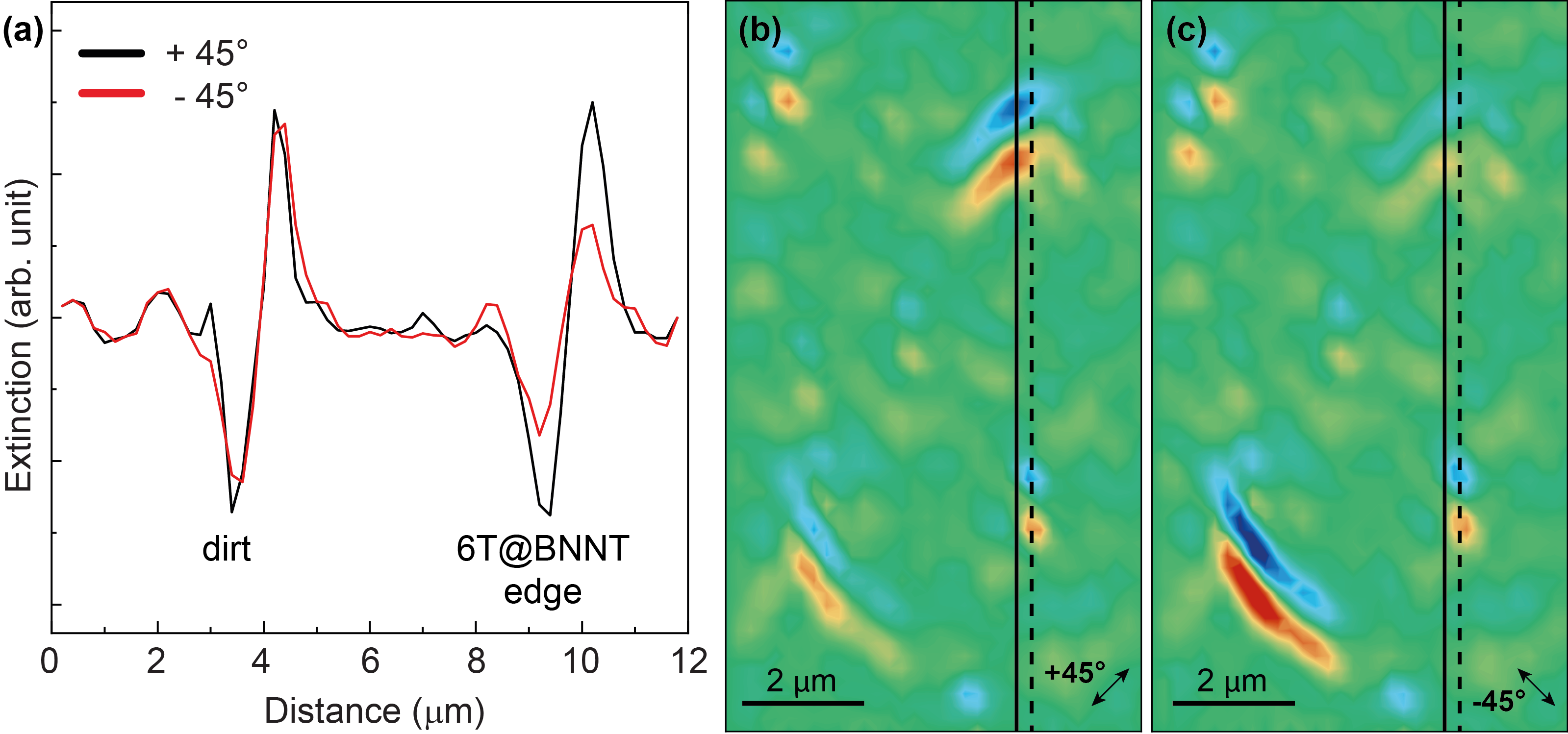}
\caption{\label{fig:SI_SMS_Map} \textbf{Polarisation dependent spatial modulation map.} \textbf{(a)} Polarisation independent line profile of dirt/non-encapsulated dye. Extinction contour plots of the studied area for \textbf{(b)} +45° and \textbf{(c)} -45° polarized light. Showing the polarisation dependence of the filled tube bundles in the upper right and lower left corner. The solid black line indicate the line profiles shown in Fig.~\ref{fig:SMS_Pol}b of the main script. The dashed line indicated the position of the line profile in \textbf{(a)}.}
\end{figure}

\newpage

\section{6T Monomer - Optical Properties}
\begin{figure}[h]
\centering
\includegraphics[width=0.5\textwidth]{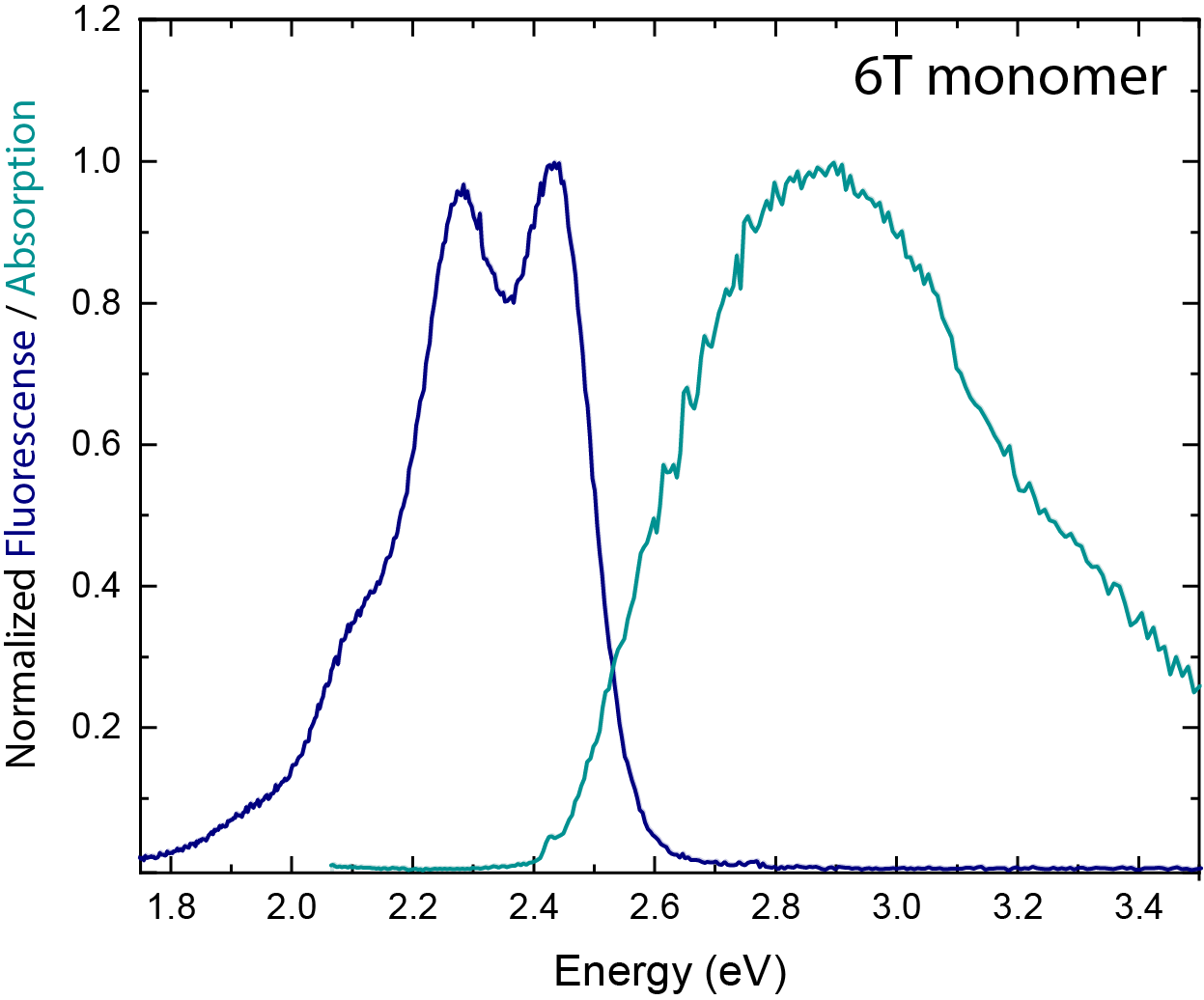}
\caption{\label{fig:Comparison} \textbf{Emission and absorption of a 6T monomer.} The dark blue (petrol blue) curve shows the fluorescence (absorption) spectrum of the 6T monomer measured in solution and excited at 425\,nm (detected at 510\,nm), showing a large Stokes shift.}
\end{figure}





\end{document}